\documentclass[runningheads]{llncs}
\usepackage[T1]{fontenc}
\usepackage{graphicx}
\usepackage{hyperref}
\usepackage{color}

\usepackage{float}
\usepackage{subcaption}
\usepackage{listings}
\usepackage[ukrainian,english]{babel}

\definecolor{gray}{rgb}{0.4,0.4,0.4}
\definecolor{darkblue}{rgb}{0.0,0.0,0.6}
\definecolor{cyan}{rgb}{0.0,0.6,0.6}
\begin{document}
\title{Unfolding the Past: A Comprehensive Deep Learning Approach to Analyzing Incunabula Pages}
\titlerunning{Unfolding the Past: A Comprehensive DL Approach to Analyzing \ldots}
\author{Klaudia Ropel\orcidID{0009-0003-1432-0718} \and
Krzysztof Kutt\orcidID{0000-0001-5453-9763} \and
Luiz~do~Valle~Miranda\orcidID{0000-0003-1838-5693} \and
Grzegorz~J.~Nalepa\orcidID{0000-0002-8182-4225}}
\authorrunning{K. Ropel et al.}
\institute{Department of Human-Centered Artificial Intelligence, Institute of Applied Computer Science, Faculty of Physics, Astronomy and Applied Computer Science, Jagiellonian University, prof. Stanis\l{}awa \L{}ojasiewicza 11, 30-348 Krak\'{o}w, Poland\\
\email{\{klaudia.ropel,krzysztof.kutt,grzegorz.j.nalepa\}@uj.edu.pl,\\luiz.dovallemiranda@doctoral.uj.edu.pl}}
\maketitle              %
\begin{abstract}
We developed a proof-of-concept method for the automatic analysis of the structure and content of incunabula pages. %
A custom dataset comprising 500 annotated pages from five different incunabula was created using resources from the Jagiellonian Digital Library. Each page was manually labeled with five predefined classes: Text, Title, Picture, Table, and Handwriting. Additionally, the publicly available DocLayNet dataset was utilized as supplementary training data. To perform object detection, YOLO11n and YOLO11s models were employed and trained using two strategies: a combined dataset (DocLayNet and the custom dataset) and the custom dataset alone. The highest performance (F1 = 0.94) was achieved by the YOLO11n model trained exclusively on the custom data. Optical character recognition was then conducted on regions classified as Text, using both Tesseract and Kraken OCR, with Tesseract demonstrating superior results. Subsequently, image classification was applied to the Picture class using a ResNet18 model, achieving an accuracy of 98.7\% across five subclasses: Decorative\_letter, Illustration, Other, Stamp, and Wrong\_detection. Furthermore, the CLIP model was utilized to generate semantic descriptions of illustrations. The results confirm the potential of machine learning in the analysis of early printed books, while emphasizing the need for further advancements in OCR performance and visual content interpretation.

\end{abstract}

\section{Introduction and Motivation}
\label{sec:intro}

The mass digitization of cultural institutions' resources, ongoing since the 1990s, has led to the creation of enormous quantities of digital copies of cultural heritage objects in digital libraries, accessible to a wide audience ranging from expert researchers to hobbyists~\cite{terras2022dh}.
However, the mere creation of digital objects and their placement on the Internet have only led to the creation of huge collections accompanied by poor and often non-standardized descriptive metadata~\cite{ranjgar2024ch}.
They usually does not go beyond the basic attributes contained in the Dublin Core standard, created in the 1990s for the purpose of indexing websites~\cite{arms2012dl}.
The lack of detailed metadata hinders the search and exploration of these resources~\cite{buwroc2022rola} and, over time, can lead to a loss of accessibility and understanding of their contents, a situation referred to as the ``digital dark age''~\cite{ghosh2015darkage}.

This is not due to a lack of appropriate standards.
The interdisciplinary community focused on cultural heritage research %
has developed or adapted several metadata schemas, thesauri, controlled vocabularies, and ontologies to suit its needs~\cite{gaitanou2024ld,ranjgar2024ch,silva2024ch,lvm2025swjournal}.
The problem is that manual creation of rich metadata is impractical due to human subjectivity and the sheer volume of digital resources~\cite{ranjgar2024ch}.
Therefore, it is worth relying on artificial intelligence (AI) methods in this task, as they support curators in creating new metadata~\cite{karterouli2021ai},
obtaining it from external sources~\cite{simou2017enriching},
or improving the quality of existing records~\cite{bobasheva2022learning}.
AI can also relieve curators of tedious tasks related to entering basic descriptions of objects~\cite{bobasheva2022learning}.
An example of this approach is the Metadata Enrichment Model (MEM), a general architecture based on linked data, computer vision models, and large language models proposed in~\cite{jig2025ijcnn}.

In this perspective, the oldest European printed materials, created between Johann Gutenberg's invention and the year 1500, known as incunabula~\cite{bolanaMirkovi2023technical} are interesting resources.
As key assets of the revolutionary period, which enabled the rapid development of science, knowledge, and economics~\cite{lacasta2022tracing,bolanaMirkovi2023technical}, they are widely studied\footnote{See~\cite{hagan2024incunabula,potten2012introduction} for some spotlights on incunabula research.}, not only by historians, but also, e.g., by researchers investigating trade and scientific networks connecting various places across Europe~\cite{kilianczyk2016introduction}.

The use of AI methods %
has the potential of supporting researchers' work on incunabula.
However, as direct successors to medieval handwritten books~\cite{kuzmenko2024incunalua}, they share some of the editorial practices with them---such as the use of non-standard typographical glyphs, word abbreviations, inconsistent word breaks between lines, and the elimination of spaces between some words~\cite{rydbergcox2010incunabula}. The presence of these elements constitutes a challenge for automatic processing, since it prevents the direct use of AI models trained on contemporary documents.

Recent research indicates that with the support of people involved in text segmentation and post-correction, it is possible to use state-of-the-art optical character recognition (OCR) tools to extract text from incunabula~\cite{reul2017case}.
However, it must be emphasized that cultural heritage objects contain a lot of important and useful information for researchers beyond the text itself -- in the case of incunabula, these include numerous decorations, engravings, handwritten notes, stamps, initials, etc.
The mere indexing of such elements would speed up the work of researchers, who would not have to browse through hundreds of pages in search of specific elements\footnote{Oral information shared during numerous discussions with Jacek Partyka, PhD, an expert on incunabula and old prints from the Jagiellonian Library, Krak\'{o}w, Poland.}.
In AI, this is the task of object detection, i.e., finding and labeling different types of objects in a drawing/photo/scan/video frame, done by models like YOLO (You Only Look Once)~\cite{redmon2016yolo}.
To the best of the authors' knowledge, object detection for digital versions of cultural heritage objects, in particular for incunabula, has not been considered in the literature.

Our goal is to fill this gap by investigating the extent to which state-of-the-art object detection models can be used to study incunabula.
Due to a lack of available datasets, the entire study began with the preparation of appropriate data using documents stored in the Old Prints Section of the Jagiellonian Library (JL)~\cite{partyka2017stare}, which served as the basis for the complete proof-of-concept demonstration of the usefulness of these models for incunabula.
In terms of its relevance to researchers, including the appropriateness of the tags/classes selected for the proof-of-concept, the study was consulted with Jacek Partyka, PhD, from the Old Prints Section of the JL. %
The remainder of the paper is structured as follows.
Sect.~\ref{sec:project} presents the experiments. %
Sect.~\ref{sec:summary} concludes the paper.

\section{Overview of the Proof-of-concept Study}
\label{sec:project}

\begin{figure}[th]
    \centering
    \includegraphics[width=0.7\textwidth]{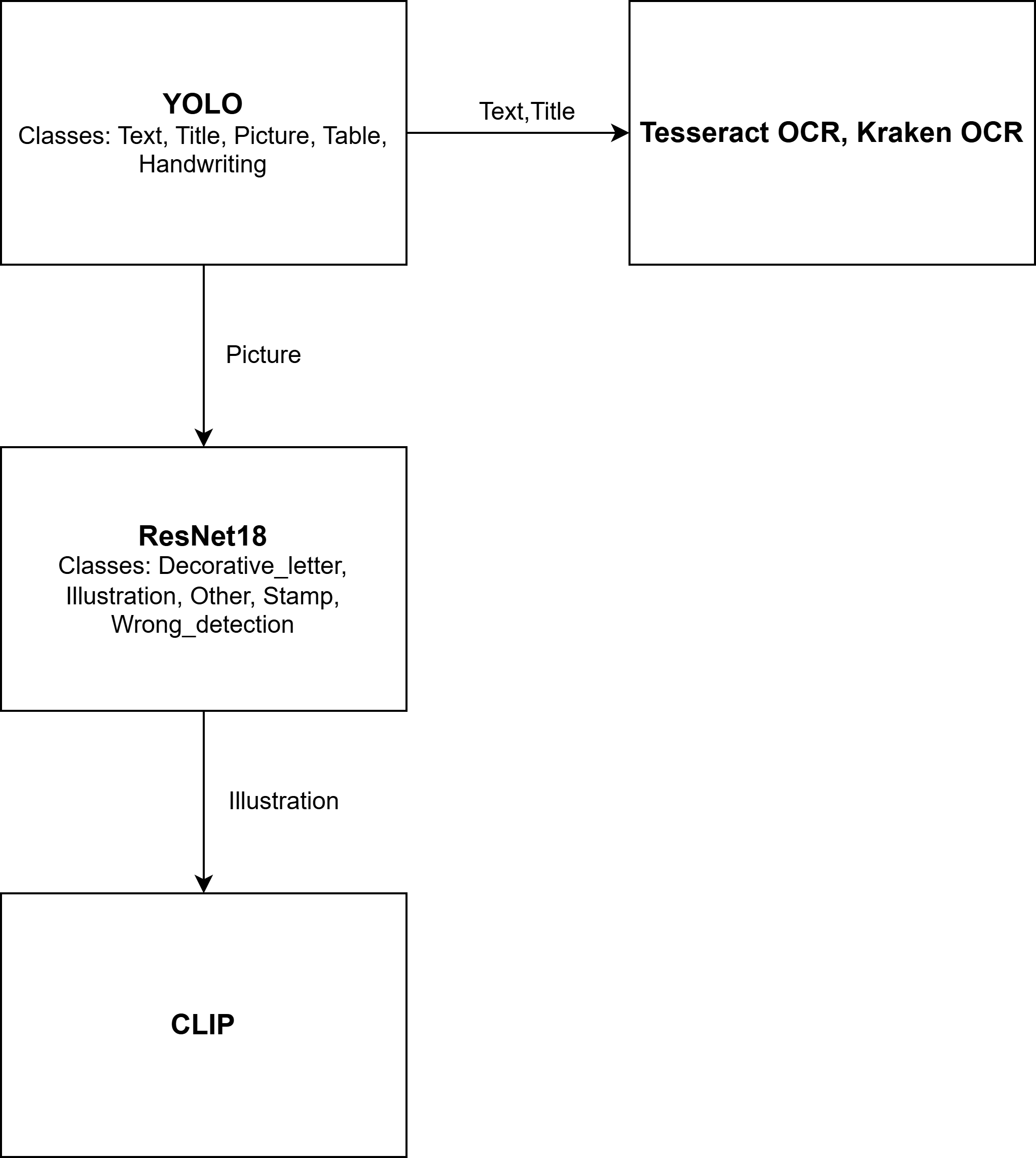}
    \caption{Processing pipeline: class detection using the YOLO model, classification of visual elements with ResNet18, and semantic interpretation of illustrations via CLIP. Textual regions (Text, Title) are processed using OCR tools.} %
    \label{fig:pipeline}
\end{figure}

The general structure of the proposed approach for analyzing incunabula pages is presented below. The diagram (see Fig.~\ref{fig:pipeline}) illustrates the main components and the data flow between models used in the project.

\begin{figure}[p]
    \centering
    \includegraphics[width=0.75\textwidth]{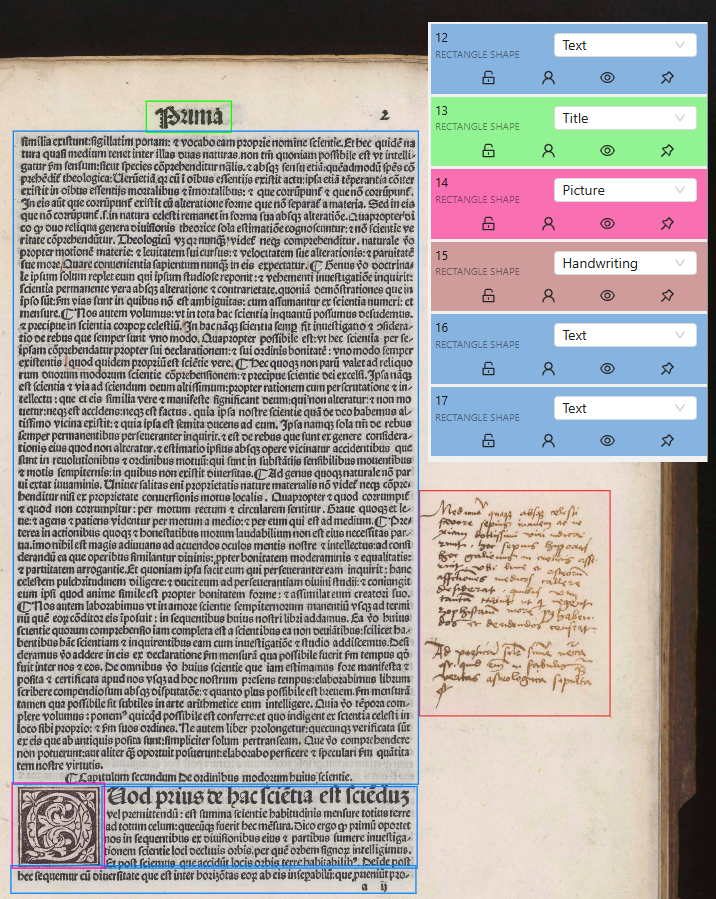}
    \caption{Example PNG file from the training set.}
    \label{fig:example_page}
\end{figure}

The first stage of the project was data collection.
After searching available sources, no existing annotated dataset of old prints/incunabula suitable for our task was found.
From the Jagiellonian Digital Library\footnote{\url{https://jbc.bj.uj.edu.pl/en/dlibra}}, five selected incunabula's scans in PDF format were retrieved and downloaded.
Using a custom script (based on the preliminary version developed for~\cite{jig2025dh}), the first 100 pages of each incunabulum were extracted and saved individually as PNG files in the corresponding subdirectories.
In total, a dataset of 500 images was created.
Each PNG file was manually annotated using five classes with the CVAT tool\footnote{\url{https://www.cvat.ai/}}.
The selected classes were \emph{Text}, \emph{Title}, \emph{Picture}, \emph{Table}, and \emph{Handwriting}.
An example page with annotations is shown in Fig.~\ref{fig:example_page}.
The resulting dataset was divided into training, validation, and test sets in proportions of 80\%, 10\%, and 10\%, respectively.

\begin{figure}[p]
    \centering
    \includegraphics[width=0.95\textwidth]{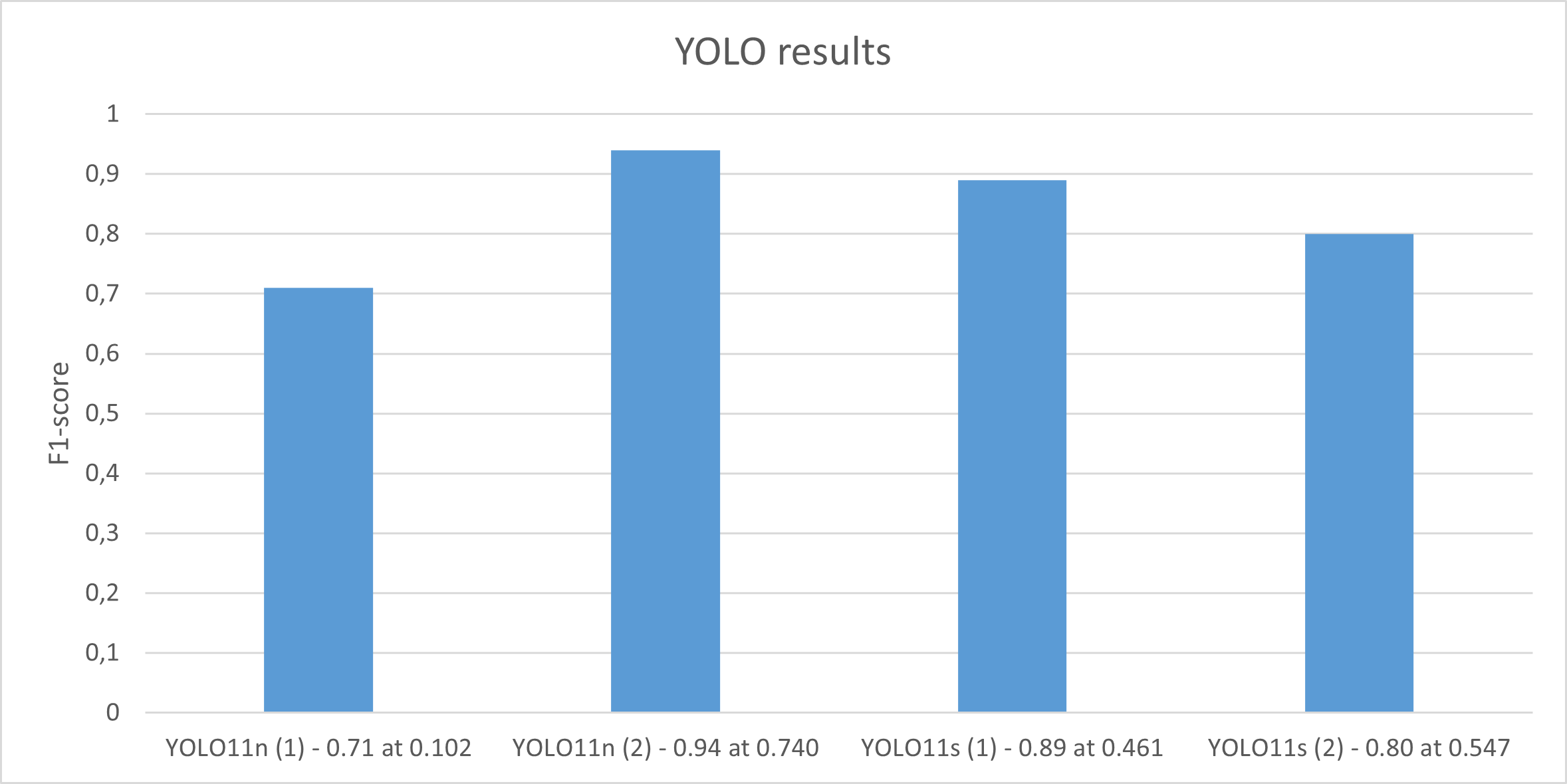}
    \caption{YOLO model results for different training datasets.}
    \label{fig:yolo_results}
\end{figure}

As an auxiliary dataset for better fine-tuning of selected models, the DocLayNet was used. DocLayNet\footnote{\url{https://github.com/DS4SD/DocLayNet}} is a large-scale, human-annotated dataset for document layout segmentation, comprising 80,863 diverse pages from six document categories, with detailed bounding-box annotations across 11 layout classes to support high-precision training and evaluation of machine learning models~\cite{pfitzmann2022doclaynet}.
DocLayNet originally contains the following classes: \emph{Caption}, \emph{Footnote}, \emph{Formula}, \emph{List-item}, \emph{Page-footer}, \emph{Page-header}, \emph{Picture}, \emph{Section-header}, \emph{Table}, \emph{Text}, and \emph{Title}.
The classes were remapped as follows:
\begin{itemize}
\item \emph{Caption}, \emph{Footnote}, \emph{List-item}, \emph{Section-header}, and \emph{Text} were merged into the \emph{Text} class,
\item \emph{Picture}, \emph{Title}, and \emph{Table} classes were retained,
\item \emph{Formula}, \emph{Page-footer}, and \emph{Page-header} classes were removed.
\end{itemize}
The DocLayNet dataset was already pre-split by the authors into training, validation, and test sets.

To perform the task of detecting individual classes (objects) in an image, two YOLO models~\cite{yolo} were used: 
YOLO11n and YOLO11s.
Both models were fine-tuned in two ways.
The first approach involved initially training the model on the DocLayNet dataset, followed by further fine-tuning on the custom 500-sample dataset.
The second approach involved training the YOLO models solely on the manually created dataset.

The metric used to compare the models was the F1-Confidence score for each of them. Both the confidence value and the F1-score range from 0 to 1, where higher values indicate greater model certainty and better prediction quality, respectively. The YOLO11n model trained using the first approach initially achieved an F1-score of 0.85 at a confidence threshold of 0.415 on the DocLayNet dataset. After further fine-tuning on the custom dataset, its F1-score dropped to 0.71 at a confidence of 0.102. The YOLO11n model trained using the second approach achieved the best average F1-score of 0.94 at a confidence threshold of 0.740. The YOLO11s model trained on both the DocLayNet and the custom dataset achieved an F1-score of 0.86 at a confidence of 0.553, and then 0.89 at a confidence of 0.461. The YOLO11s model trained solely on the custom dataset reached an F1-score of 0.80 at a confidence of 0.547. Fig.~\ref{fig:yolo_results} presents a summary of the final results obtained for the selected YOLO models. A number in parentheses indicates whether the first (1) or second (2) training approach was applied.

The best results in terms of both F1-score and confidence threshold were achieved by the YOLO11n model trained solely on the custom dataset. Therefore, it was selected as the base model for classifying the chosen classes on incunabulum pages.

The next stage of the project involved the goal of automatically recognizing text from the Text and Title classes, i.e., performing OCR. The best publicly available OCR tools suitable for old prints were selected for testing: Kraken OCR\footnote{\url{https://github.com/mittagessen/kraken}} and Tesseract OCR\footnote{\url{https://github.com/tesseract-ocr/tesseract}}. To perform OCR, the previously trained YOLO11n model was used to extract and save the areas labeled as \emph{Text} and \emph{Title} into separate directories. Tesseract OCR performed significantly better than Kraken OCR. However, the results still included many errors and misread characters. The main issue was the lack of suitable training data (transcribed incunabula), which could have been used to fine-tune the model. In that case, the performance of the OCR would likely be improved.

\begin{figure}[t]
    \centering

    \begin{subfigure}[b]{0.3\textwidth}
        \centering
        \includegraphics[width=\textwidth]{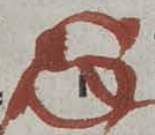}
        \caption{Example of the \emph{Decorative letter} class.}
        \label{fig:decorative_letter}
    \end{subfigure}
    \hfill
    \begin{subfigure}[b]{0.3\textwidth}
        \centering
        \includegraphics[width=\textwidth]{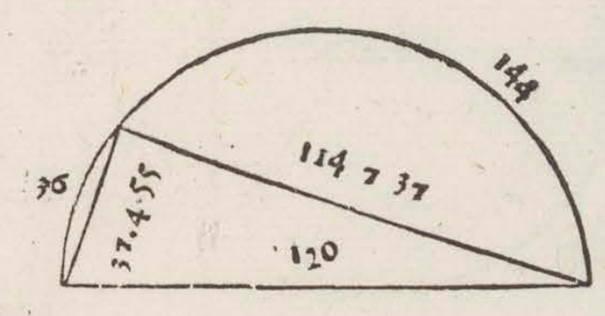}
        \caption{Example of the \emph{Illustration} class.}
        \label{fig:illustration}
    \end{subfigure}
    \hfill
    \begin{subfigure}[b]{0.3\textwidth}
        \centering
        \includegraphics[width=\textwidth]{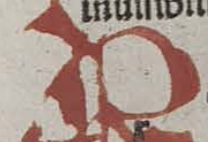}
        \caption{Example of the \emph{Wrong detection} class: a poorly cropped decorative letter.}
        \label{fig:wrong_detection}
    \end{subfigure}

    \vspace{0.5em}

    \hfill
    \begin{subfigure}[b]{0.25\textwidth}
        \centering
        \includegraphics[width=\textwidth]{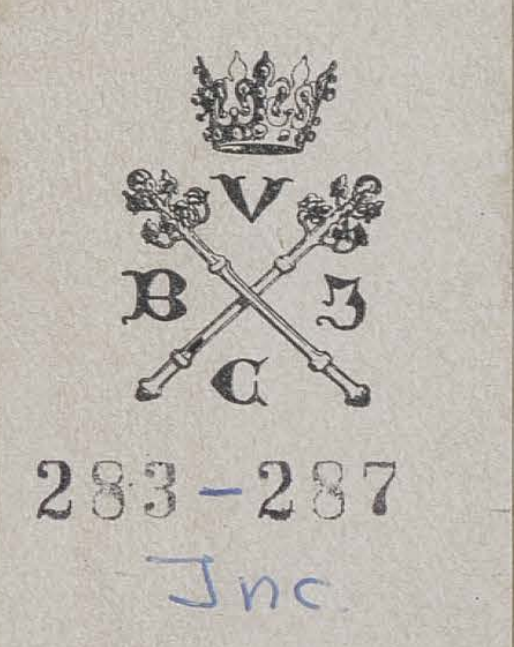}
        \caption{Example of the \emph{Stamp} class.}
        \label{fig:stamp}
    \end{subfigure}
    \hfill
    \begin{subfigure}[b]{0.25\textwidth}
        \centering
        \includegraphics[width=\textwidth]{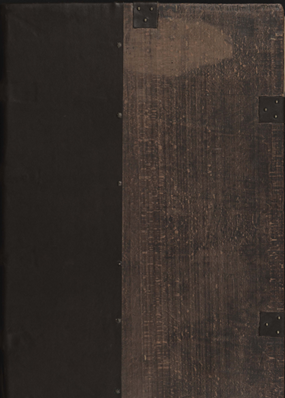}
        \caption{Example of the \emph{Other} class.}
        \label{fig:other}
    \end{subfigure}
    \hfill
~
    \caption{Examples of classified image elements from the Picture class.}
    \label{fig:picture_examples}
\end{figure}

The final stage involved additional classification of the previously labeled \emph{Picture} class.
The previously trained YOLO11n model was once again used to extract and save the \emph{Picture} class regions from the selected incunabula into separate directories.
A ResNet18 model was then trained to classify five categories:
\begin{itemize}
    \item \emph{Decorative\_letter}, which included ornamental letters (Fig.~\ref{fig:decorative_letter}),
    \item \emph{Illustration}, which covered various illustrations and, for convenience during training, also included initials with a large number of decorative elements (Fig.~\ref{fig:illustration}),
    \item \emph{Other}, such as elements of the book cover (Fig.~\ref{fig:other}),
    \item \emph{Stamp}, referring to detected stamps (Fig.~\ref{fig:stamp})
    \item \emph{Wrong\_detection}, which included elements that were misclassified or poorly cropped by YOLO11n (Fig.~\ref{fig:wrong_detection}).
\end{itemize}
The ResNet18 model achieved an accuracy of 98.70\%. Examples of these categories are shown in Fig.~\ref{fig:picture_examples}.

For the \emph{Illustration} class, it was decided to apply the CLIP model.
CLIP\footnote{\url{https://github.com/openai/CLIP}} is an AI model that learns to associate images with natural language descriptions.
Thanks to this, it can recognize what is depicted in an image, even if it has not been specifically trained for that particular task~\cite{radford202clip}.
The following example descriptions were selected for the images:
\begin{lstlisting}[language=Python]
descriptions = [
    "religious scene",
    "astrology or alchemy",
    "medieval engraving",
    "Renaissance-style illustration",
    "magic and divination",
    "printer's ornament",
    "decorative book illustration",
    "mathematics"
]
\end{lstlisting}
For example, for Fig.~\ref{fig:illustration}, the CLIP model selected the description "mathematics". To generate more accurate and meaningful descriptions, the involvement of a person more familiar with the illustrations found in the incunabula would be helpful.

Training a model to recognize specific elements of the \emph{Decorative\_letter} class would be too great a challenge, as there is no single rule for how ornamental letters appeared across all incunabula.
A single letter can look completely different in each book.
The \emph{Handwriting} class in the selected incunabula was too limited and difficult for us to decipher in order to train or use any model specializing in handwritten text.
In the future, OCR models could be used to recognize text in the \emph{Table} class to allow for a better interpretation of the content of a given incunabulum.

\section{Discussion and Future Works}
\label{sec:summary}

The conducted experiments allow for the formulation of several key conclusions. First, the use of an external dataset (DocLayNet) in the training process of detection models did not improve performance in the task of classifying elements within incunabula pages. While it contributed to greater generalization capabilities of the models, this was not aligned with the main objective of the project, which was the precise analysis of specific early printed books. Due to hardware limitations, it was not possible to fine-tune larger YOLO model variants, which restricted the range of tested architectures. In the area of text recognition (OCR), a significant barrier was encountered — the lack of appropriate training data tailored to the specificity of incunabula. As a result, the quality of text recognition was unsatisfactory. On the other hand, the application of the ResNet18 model for the classification of graphic subcategories (Picture) yielded very promising results. The model achieved high accuracy and confirmed its usefulness in visual content analysis of historical documents.

To improve future results, one of the first steps could involve preprocessing the scanned pages, including enhancing image resolution, contrast, and color balance. The application of binarization or adaptive thresholding techniques may further improve text legibility, potentially leading to better OCR outcomes. Another important development direction involves expanding the custom dataset, both in terms of the number of examples and the diversity of historical sources, which would enable the training of more complex models without compromising their performance on the target data. Further progress in text recognition (OCR) and semantic content interpretation will require the creation of specialized, domain-adapted training datasets, particularly containing transcribed texts and detailed annotations of illustrations from incunabula. In the longer term, the use of techniques such as reinforcement learning may also be considered.

\begin{credits}
\subsubsection{\ackname}
We would like to express our sincere thanks to Jacek Partyka from the Jagiellonian Library, who patiently introduced us to the meanders of research on %
incunabula.
This publication was funded by a flagship project ``CHExRISH: Cultural Heritage Exploration and Retrieval with Intelligent Systems at Jagiellonian University'' under the Strategic Programme Excellence Initiative at Jagiellonian University.
The research has been supported by a grant from the Priority Research Area (DigiWorld) under the Strategic Programme Excellence Initiative at Jagiellonian University.
We gratefully acknowledge Polish high-performance computing infrastructure PLGrid (HPC Center: ACK Cyfronet AGH) for providing computer facilities and support within computational grant no. PLG/2025/018037.
This work benefited from the use of language models to support proofreading and enhance readability.

\subsubsection{\discintname}
The authors have no competing interests to declare that are
relevant to the content of this article.
\end{credits}

\bibliographystyle{splncs04}
\bibliography{geistbib/culheripub,geistbib/culheriteam,ukrainian}

\end{document}